# Scalable quantum register based on coupled electron spins in a room temperature solid

P. Neumann[1]*, R. Kolesov[1]*, B. Naydenov[1], J. Beck[1], F. Rempp[1], M. Steiner[1], V. Jacques[1], G. Balasubramanian[1], M.L. Markham[2], D.J. Twitchen[2], S. Pezzagna[3], J. Meijer[3], J. Twamley[4], F. Jelezko[1†] and J. Wrachtrup[1†]

[1] 3. Physikalisches Institut, Universität Stuttgart, Stuttgart, 70550, Germany

[2] Element Six Ltd., King's Ride Park, Ascot, Berkshire SL5 8BP, UK

[3] Central laboratory of ion beam and radionuclides, Ruhr Universität Bochum, 44780, Germany

[4] Centre for Quantum Computer Technology, Macquarie University, Sydney 2109, Australia

* These authors contributed equally to this work

[†] e-mail: F.J. (f.jelezko@physik.uni-stuttgart.de) or J.W. (wrachtrup@physik.uni-stuttgart.de)

**Realization of devices based on quantum laws might lead to building processors that outperform their classical analogues and establishing unconditionally secure communication protocols[1]. Solids do usually present a serious challenge to quantum coherence. However, owing to their spin-free lattice and low spin orbit coupling, carbon materials[2,3] and particularly diamond[4-6] are suitable for hosting robust solid state quantum registers. We show that scalable quantum logic elements can be realized by exploring long range magnetic dipolar coupling between individually addressable single electron spins associated with separate color centers in diamond. Strong distance dependence of coupling was used to characterize the separation of single qubits 98 Å with unprecedented accuracy (3 Å) close to a crystal lattice spacing. Our demonstration of coherent control over both electron spins, conditional dynamics, selective readout as well as switchable interaction, opens the way towards a room temperature solid state scalable quantum register. Since both electron spins are optically addressable, this solid state quantum device operating at ambient conditions provides a degree of control that is currently available only for atomic systems.**

One of the greatest challenges in quantum information technology is to build a room temperature scalable quantum processor[7]. Isolated electron and nuclear spins in solids are considered to be among the most promising candidates for qubits in that respect[3,8]. Several benchmark experiments including entanglement and elements of quantum memory[9] have been achieved with spin ensembles, but ultimate functionality requires encoding quantum information into single spins. This however creates serious challenges in readout, addressing and nano-engineering single spin arrays. The availability of photon assisted single spin readout[10,11] and the possibility to create single defects by ion implantation[12,13] make nitrogen-vacancy (NV) defects in diamond one of the most promising candidates in this respect. Paramagnetic nuclei in the vicinity of the electron spin can be used as auxiliary qubits with even more favourable relaxation properties[14]. As a consequence coherence between electron and nuclear spin qubits has been exploited for showing all basic elements of a room temperature quantum register[5,15-17]. The size of these registers however is limited to a few quantum bits owing to the limited number of nuclear spins which can be addressed in frequency space[15,18]. A critical step in scalability is to develop a technique allowing mutual coupling of individual optically addressable quantum systems.

The system used in this study is a pair of single electron spins associated with separate NV defects in diamond. A single defect consists of a substitutional nitrogen atom in the diamond lattice and an adjacent vacancy (Fig. 1a, b). The electron spin triplet ground state of the defect shows a spin-dependent fluorescence emission enabling measurements of the spin state of single dopants using highly sensitive fluorescence microscopy techniques[10,14]. Recent experiments showing that NV defects can be coupled to spins of substitutional nitrogen has highlighted avenues for extending the scalability potential of the diamond quantum register[6,19-21]. In general, magnetic dipolar coupling can be used as a mechanism allowing the generation of entanglement in cases where coupling prevails over decoherence. The spin Hamiltonian describing the dipole-dipole interaction between two coupled spins is

$$H_{dip} = \frac{\mu_0 g^2 \mu_B^2}{4\pi r^3}\left[\hat{S}_{NV\,A}\cdot\hat{S}_{NV\,B} - 3\left(\hat{S}_{NV\,A}\cdot\vec{r}\right)\left(\hat{S}_{NV\,B}\cdot\vec{r}\right)\right] \tag{1}$$

where $\mu_0$ is the magnetic permeability, $g$ is the electronic $g$-factor, $\mu_B$ is Bohr magneton, $r$ is the distance between spins, and $\vec{r}$ the unit vector connecting them. The last term defines the angular dependence of coupling between NV defects and the prefactor is about 70 kHz for $r$=10 nm (these values are relevant to experiments presented in this letter). Such a distance between two colour centres allows their individual addressing using modern nonlinear microscopy techniques such as stimulated emission depletion microscopy (STED)[22]. At the same time coupling within a few tens of kHz range might appear to be too weak to be detectable owing to decoherence which is always present in solids. However in diamond material with a reduced concentration of paramagnetic $^{13}C$ spins, coherence times can reach milliseconds at room temperature[4]. This letter presents the first step towards the realization of an optically addressable quantum register by demonstrating coherent coupling between two single NV defects engineered in such an ultrapure crystal.

Colour centre pairs were created by implanting patterns of high energy nitrogen ions within small spots into synthetic isotopically enriched $^{12}C$ diamond . Figure 1c shows a pattern of implanted defects having 2 defects per spot on average. The implantation sites were investigated by measuring the fluorescence autocorrelation function $g^2(\tau)$ and ground state depletion microscopy[22] selecting a pair of colour centres with distance less than 30 nanometres. Since individual addressing by STED exceeds our current experimental capabilities, fluorescence lifetime imaging (FLIM) analysis was employed to resolve the two defects optically. In this approach we use a contrast mechanism based on different decay rates of the excited state spin sublevels[23] allowing to unravel the individual point spread functions corresponding to the emission of two colour centres . Figure 1d shows the image of two resolved defects with the lateral distance of 8±3 nm. At such short distances the dipolar interaction between these two electron spins is expected to be observable in the magnetic resonance spectral response of the spins. Owing to the orientation dependence of the Zeeman energy, spin transitions of the two defects have different energies when colour centres have different orientation with respect to the field, thus allowing their individual addressing (Fig. 2a). The electron spin resonance (ESR) spectrum can be described by the effective ground state spin Hamiltonian

$$H = \sum_{i=NV\,A}^{NV\,B} \left( \hat{S}_i \underline{\underline{D}} \hat{S}_i + \mu_B g_e \underline{B} \hat{S}_i \right) + H_{dip}. \qquad (2)$$

Here $\underline{\underline{D}}_{A/B}$ are the zero-field splitting tensors which mainly split $m_S^{A,B} = 0 \left( |0\rangle_{A,B} \right)$ and $m_S^{A,B} = \pm 1 \left( |\pm 1\rangle_{A,B} \right)$ levels by roughly $D$=2.87 GHz and which are differently oriented for the two centres. The Zeeman energy of the electron spins is expressed by the second term in the sum with the magnetic field $\underline{B}$. The last term explicitly given in Equation 1 describes the magnetic coupling between spins.

The most appropriate way to detect the static magnetic dipolar interaction is through spin selective Ramsey experiments. To this end one of the two electron spins (spin *A*), is brought into a coherent superposition of its two eigenstates $(|0\rangle_A + |1\rangle_A)/\sqrt{2}$. This superposition acts as a sensitive magnetometer allowing us to resolve the magnetic field created by the second distant spin (spin *B*) with $\omega_A^{Ramsey} = \omega_A^{Larmor} + \omega_A^{dip} \cdot m_S^B$. Here $\omega_A^{\mathrm{Lamor}}$ is the Larmor frequency of spin *A* and $\omega^{\mathrm{dip}}$ is the dipolar coupling frequency of the spin pair. In the Fourier transformation analysis of the Ramsey fringes (Fig. 2b) three different lines of spin $A$ are visible which are related to three different eigenstates of spin $B$ $|0,\pm1\rangle_B$ which we prepared by transition selective microwave pulses. Besides showing coupling among qubits and their control, the initialization of the quantum register is of prime importance. Since the relative intensity of coupling induced multiplets is related to the accuracy by which either of the two spins is brought into a specific starting state, an estimation of the degree of efficiency of the initialization process can be made. We estimate that with a probability of 88±4% spin *B* is initialized into its $|0\rangle$ state (higher than previously reported value for ensembles[24]). Note that our result is just a lower bound because it contains errors, e.g. from imperfect $\pi$ pulses, which lower the estimated value.

To resolve the relative positions of the two centres more accurately it is best to use echo based techniques that exploit the long phase memory time $T_2$. Given the capability to separately address the two electron spins, a prominent way to measure the interaction between electron spins is based on a state dependent phase shift method similar to double electron-electron spin resonance used in electron spin resonance spectroscopy. In such a pulse sequence a two pulse echo measurement is performed on spin *A* whereas spin *B* is flipped. The pulse sequence is designed such that all components of the $\underline{B}$-field acting on spin *A,* which are static during the pulse spacing time, are refocused except the field originating from flipped spin *B*. The associated phase is $\Delta\varphi = \gamma \cdot \delta B \cdot \mathrm{T}$ where $\delta B$ is the change in magnetic field at spin *A* induced by spin *B* and $\gamma$ the gyromagnetic ratio. Hence coupling causes a modulation of the signal as a function of time instant T when the spin *B* is flipped (Fig. 3a). This technique can be employed for precise characterization of the relative position of the two spins. Result is displayed in Figure 3c where we have placed the first spin at the origin of coordinates system (0|0|0) (red dot), and the *x*-*y*-plane is parallel to the diamond surface. The location of the second spin is measured to be within the red ellipsoidal region (Fig. 3c, upper part), whose size is defined by the experimental accuracy. Only six possible lattice positions fall within this area (Fig. 3c, lower part). The two qubits are measured to have a distance of $r$=9.8±0.3 nm and a lateral separation with respect to the diamond surface of 8.8±0.3 nm which is in nice agreement with the optical image analysis.

Our experiments show that two qubits can be addressed and readout separately using transition selective microwave fields. Moreover spin $A$ and $B$ can be coherently controlled. A final element necessary to complete the toolbox of quantum operations in our register is conditional spin dynamics among spin $A$ and $B$. The spin flip executed on spin $B$ in the experiment presented in Figure 3a is transition dependent, therefore the magnitude of the echo signal of spin $A$ depends on the state of spin $B$. The fidelity of such conditional spin flip is maximized after $\Delta\varphi=\pi$. Experimentally the observed fidelity is limited by the coupling strength of the two spins (≈40 kHz) and the phase memory time of spin $A$ (110 μs). Moreover, since the nitrogen-vacancy centre has spin $S$=1, the gate operation can be improved by using the higher magnetic moment associated with a forbidden transition. If spin $B$ is in state $|-1\rangle_B$ at the beginning of the sequence and is then brought from $|-1\rangle_B$ to $|+1\rangle_B$ a doubled oscillation frequency is expected. Such forbidden spin transitions were driven by composite microwave pulses. Furthermore, when the echo is performed on the transition $|-1\rangle_A \leftrightarrow |+1\rangle_A$ a four times faster oscillation occurs (Fig. 3b). We conclude that using double quantum coherences can considerably improve the speed and fidelity of quantum logic gates.

The coherent coupling between spins can be used for the creation of non-classical correlation in a two-qubit quantum register. The quantum circuit of this operation (Fig. 4a) employs controlled single qubits operations realized by Hadamard gates and two-qubit phase shift gate. The observed signal oscillates between a maximum value corresponding to the quantum state $(i|0,-1\rangle - |0,0\rangle)/\sqrt{2}$ at $\tau = (0,2,4,\ldots)\pi/(\gamma \cdot \delta B)$ and minimum value for the entangled state $|\Phi\rangle$ at $\tau = (1,3,5,\ldots)\pi/(\gamma \cdot \delta B)$ when the evolution period τ is varied (Fig. 4b). If the starting state is $|-1,0\rangle$ the oscillation is between states $(i|-1,-1\rangle - |-1,0\rangle)/\sqrt{2}$ and $(|-1,0\rangle + i|0,-1\rangle)/\sqrt{2}$, the latter one being the entangled state $|\Psi\rangle$. It is important to mention that the above described state preparation is computed for the ideal case (without taking into account decoherence). In our experiments, owing to the surprisingly short $T_2$ of $NV\,B$ (2 μs), a random phase ϕ is acquired which is not accessible and no proper quantum state tomography can be performed. The fast decoherence of one of the spins qubits exceeds the coupling strength between the spins by an order of magnitude; hence magnetic coupling itself cannot be responsible for the dephasing.

Such fast decoherence limits the generation of entanglement and in our case we have not yet reached a regime which allow us to test Bell`s inequalities. While values for $T_2$ in excess of 2 ms have been recorded for single NV centres in such isotopically enriched CVD diamond, the reduced values reported here for centres created using high energy implants (See Methods) motivates further material science research for detailed understanding and optimization of diamond lattice dynamics in the presence of two closely spaced colour centres.

Results presented in this letter show a significant step towards realization of a scalable room temperature solid state quantum information processing device. By artificially creating nitrogen-vacancy colour centres in an ultrapure and isotopically engineered CVD grown diamond we observe coupling of two single optically addressable qubits. Entanglement purification protocols[25] using carbon nuclear spins as auxiliary "storage" qubits can be explored for achieving high fidelities of entanglement. Advanced single atom fabrication technologies might further improve the engineering accuracy of the quantum register[26].Most notably the coupling occurred over such long distances that state-of-the-art ultrahigh resolution optical microscopy will allow the spins to be addressed separately. This will result in a highly versatile room temperature quantum device.

## Methods summary

The diamond sample is an ultrapure type IIa synthetic crystal produced by microwave plasma assisted chemical vapour deposition. The concentration of $^{12}C$ has been increased to 99.99% by using isotopically enriched methane in the growth process. NV centres have been created by implantation of $^{14}N$ ions with energy of 18 MeV and subsequent annealing at 800°C for 2 hours. For such high implantation energy the nitrogen to NV conversion efficiency is better than previously reported owing to larger number of vacancies created (average NV creation efficiency 21%). In order to increase the probability to create pairs of NV centres at small separations, several arrays of implantation sites with varying numbers of ions per spot have been created.

NV centre pair candidates were detected by a home-build confocal microscope able to record optically detected magnetic resonance and the 2$^{nd}$ order correlation function. High resolution images were obtained by structured illumination microscopy and fluorescence lifetime imaging techniques in order to ensure that the distance of the NV centres allows for observation of magnetic dipole-dipole coupling. All measurements have been carried out at room temperature.

Determining the relative position of two separate spins relies on the change of their coupling strength depending on the orientation of their quantisation axes (Equation (1)). These axes are changed by rotating the magnetic field. Eventually, recording the coupling frequencies for various magnetic field settings allows determining the appropriate relative position of the two centres. See supplementary online material for further information.

**Supplementary Information** is linked to the online version of the paper at www.nature.com/naturephysics

**Acknowledgements** This work was supported by the EU (QAP, EQUIND, NEDQIT), DFG (SFB/TR21 and FOR730), NIH, Landesstiftung BW and the Volkswagen Stiftung.

**Author Contributions** P.N., R.K., M.S., J.B., V.J., B.N. and F.J. performed the experiments;. D.J.T. and M.L.M., designed and performed synthesis of diamond material, J.T. and F.R. performed theory analysis of dipolar coupling of NV defects, J.M. B.N. and S.P performed implantation of single NV defects; P.N., J.W. and F.J. wrote the paper. All authors discussed the results, analysed the data and commented on the manuscript.

**Author Information** Correspondence and requests for materials should be addressed to F.J. (f.jelezko@physik.uni-stuttgart.de) or J.W. (wrachtrup@physik.uni-stuttgart.de).

## Figure Legends

**Figure 1. Optical microscopy of coupled defects in diamond. a,** (lower part) Illustration of two coupled NV defect centres in diamond manipulated by green laser light (532 nm) and microwave radiation. (upper part) Structure of the NV defect within the diamond lattice showing nitrogen (N) and carbon (C) atoms and the vacancy (V). **b,** Energy level scheme of the NV colour centre comprises ground (GS) and excited (ES) state triplets and a metastable singlet state (MS). Excitation light is shown as green arrows and fluorescence as red arrow. Spin-dependent intersystem-crossing rates (black arrows) lead to spin polarization into the $m_S = 0\ (|0\rangle)$ level and lead to a different fluorescence level for $|0\rangle$. **c,** Fluorescence microscopy image of the pattern of defects created by high energy implantation of nitrogen. The close-up shows the yet unresolved NV centre pair. **d,** Spot containing two coupled colour centres resolved by FLIM. The two sub-images for the two centres with different fluorescence lifetime (11ns, 7 ns) are correlated. The result reveals a displacement of the two centres of roughly 10 nm.

**Figure 2. Dipolar coupling between two single spins. a,** Optically detected electron spin resonance spectra of two coupled defects show fluorescence intensity $I_{PL}$ depending on the microwave frequency. Owing to different orientations in the lattice, single spins are addressable using spin selective frequency pulses. The corresponding electron spin levels of the ground state of the NV centre are shown for both centres. **b,** Shift of spectral line of spin *A* depending on state of spin *B* mediated by magnetic coupling. Graph shows the Fourier transform of the Ramsey fringes of spin *A* after preparing spin *B* in one of its eigenstates $(|-1\rangle, |0\rangle, |+1\rangle)$ by application of transition selective microwave $\pi$ pulses. The spectral region shown is that of a single hyperfine line. Each line is fitted by three Gaussians. The combined energy level diagram shows the respective transitions.

**Figure 3. Unravelling the structure of quantum register a,** Double electron-electron spin resonance using microwave pulses selective between NV *A* and *B*. Echo modulation is observed for spin *A* when the coupled spin *B* is flipped within the echo sequence. Modulation

amplitude frequency corresponds to dipolar coupling strength (42 kHz)). Control experiment with detuned driving field for spin *B* is also shown (grey line). The amplitude of $I_{PL}$ is normalized to the signal of a spin flip from $|0\rangle$ to $|\pm 1\rangle$. **b,** Fast quantum gate operation using double quantum coherence transition of spin *B*. (red line) and both spins (gray line). **c,** Reconstruction of two-qubit register geometry. Relative position of one spin with respect to the other positioned at the origin of coordinate system for convenience (upper part). The red area which includes six possible lattice locations of the second spin is defined based on spin resonance measurements. Magnification of uncertainty area is shown in the lower part. Lattice sites are displayed as grey spheres and those inside the red area are bigger and black.

**Figure 4. Exploring dipolar coupling for creating an entangled state. a,** Pulse sequence for generation of entanglement in this two qubit quantum register (|Φ⟩-state in presented case). Preparing both centres in the state $|00\rangle_{A,B}$, by optical pumping, an echo is performed on spin *A*. Right after the waiting time $\tau = \pi/(\gamma \cdot \delta B)$ and the echo's π pulse on spin *A* a π/2 pulse on NV *B* is applied resulting in a state $(|-1,-1\rangle + i|-1,0\rangle + |0,-1\rangle + i|0,0\rangle)/2$. This is turned into $(-i|-1,-1\rangle + i|-1,0\rangle + |0,-1\rangle + |0,0\rangle)/2$ after the second waiting time τ. Finally a π/2 pulse on spin *A* produces the quantum state $(i|-1,-1\rangle - |0,0\rangle)/\sqrt{2}$ which is the maximally entangled |Φ⟩ Bell state. **b,** Fluorescence intensity depending on pulse delay τ in entanglement sequence (red and black data points). The minima for the |Φ⟩ (black) and the maxima for the |Ψ⟩ states (red) mark the τ values of highest fidelity $\tau = (1,3,5,\dots)\pi/(\gamma \cdot \delta B)$. The red and black curves are calculations using the coupling between the centres and their coherence times. In the same way the fidelity is simulated (gray curve).

**Figure 1,** P. Neumann et al.

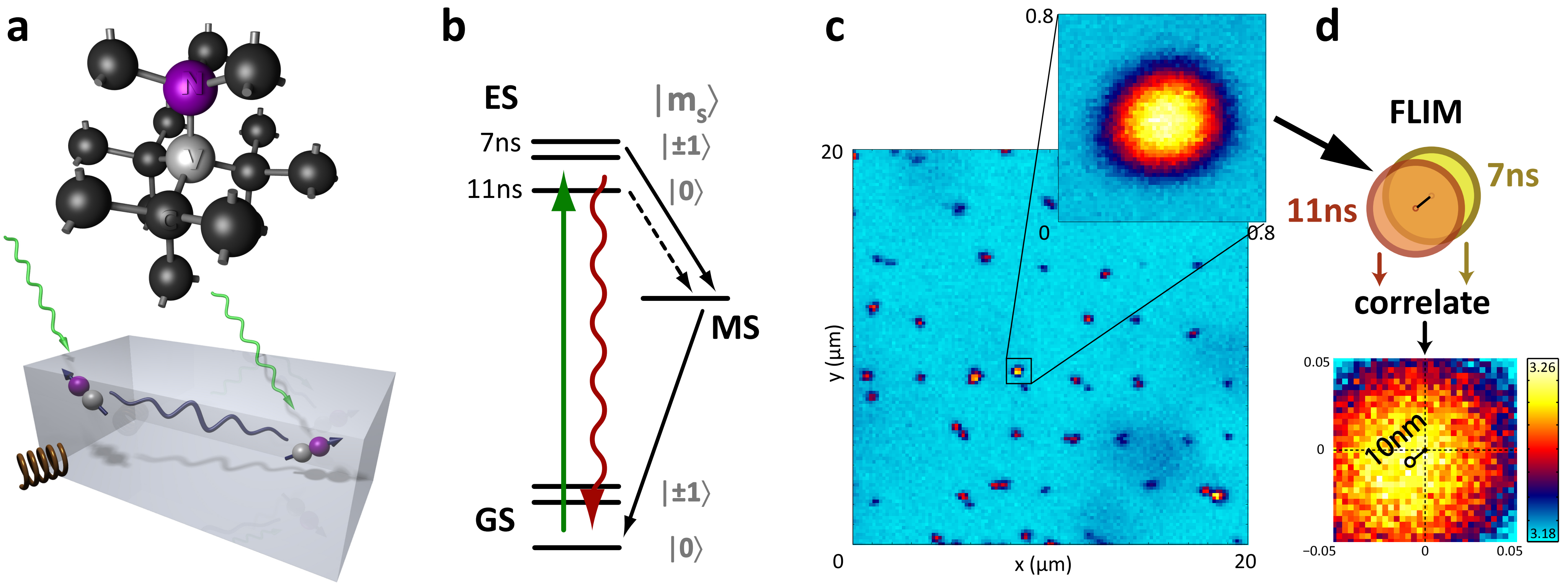

# **Figure 2,** P. Neumann et al.

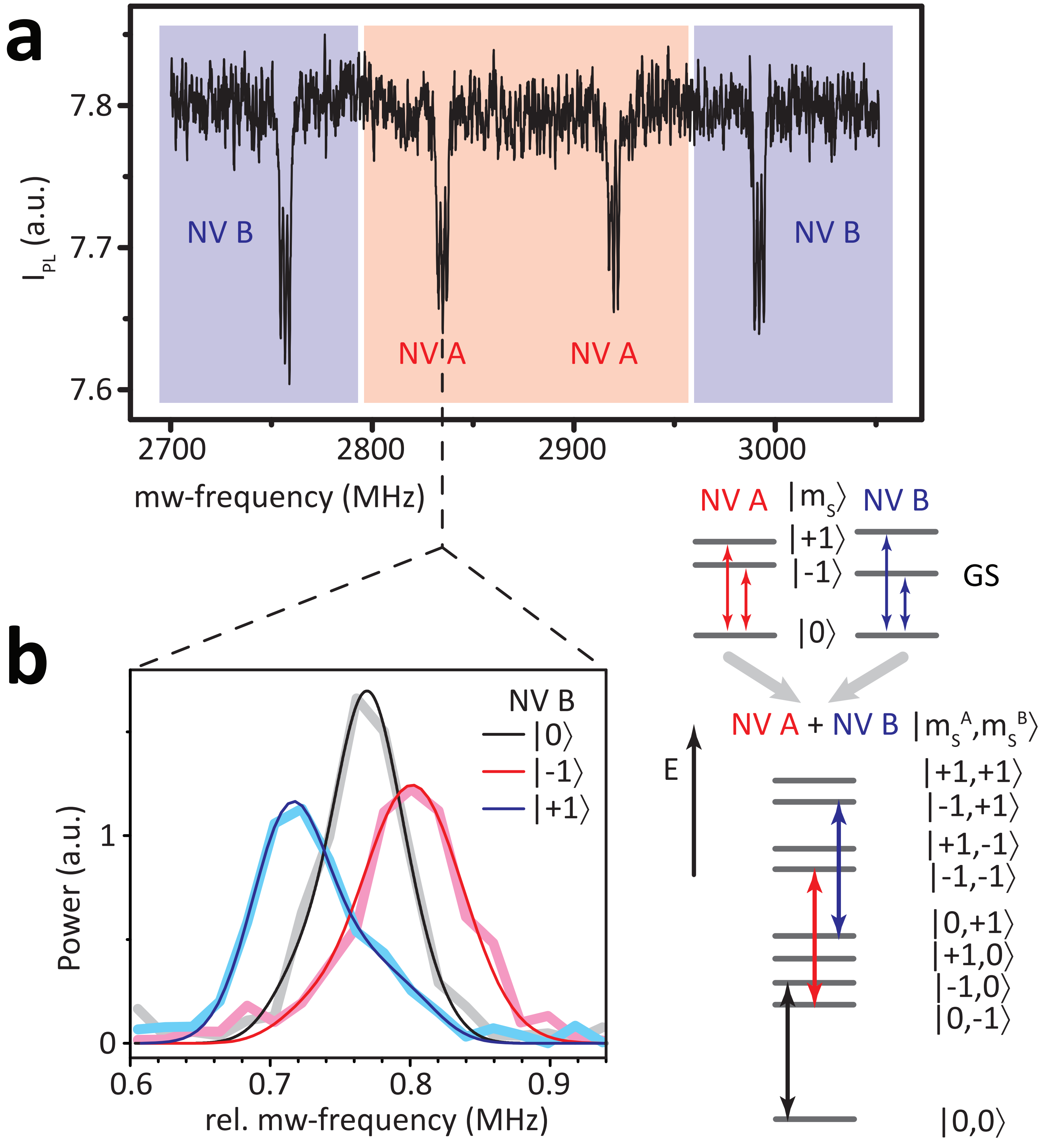

**Figure 3,** P. Neumann et al.

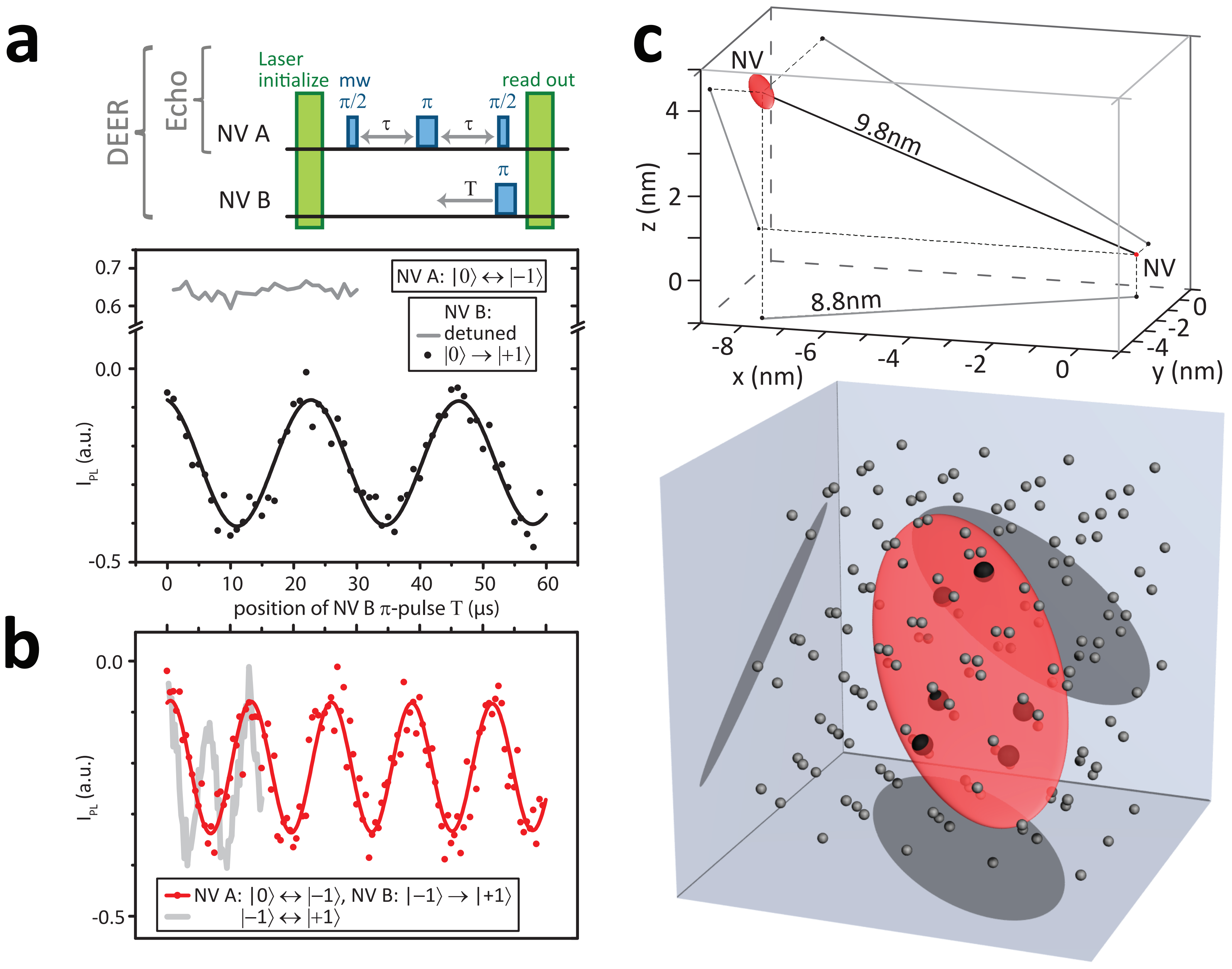

# Figure 4, P. Neumann et al.

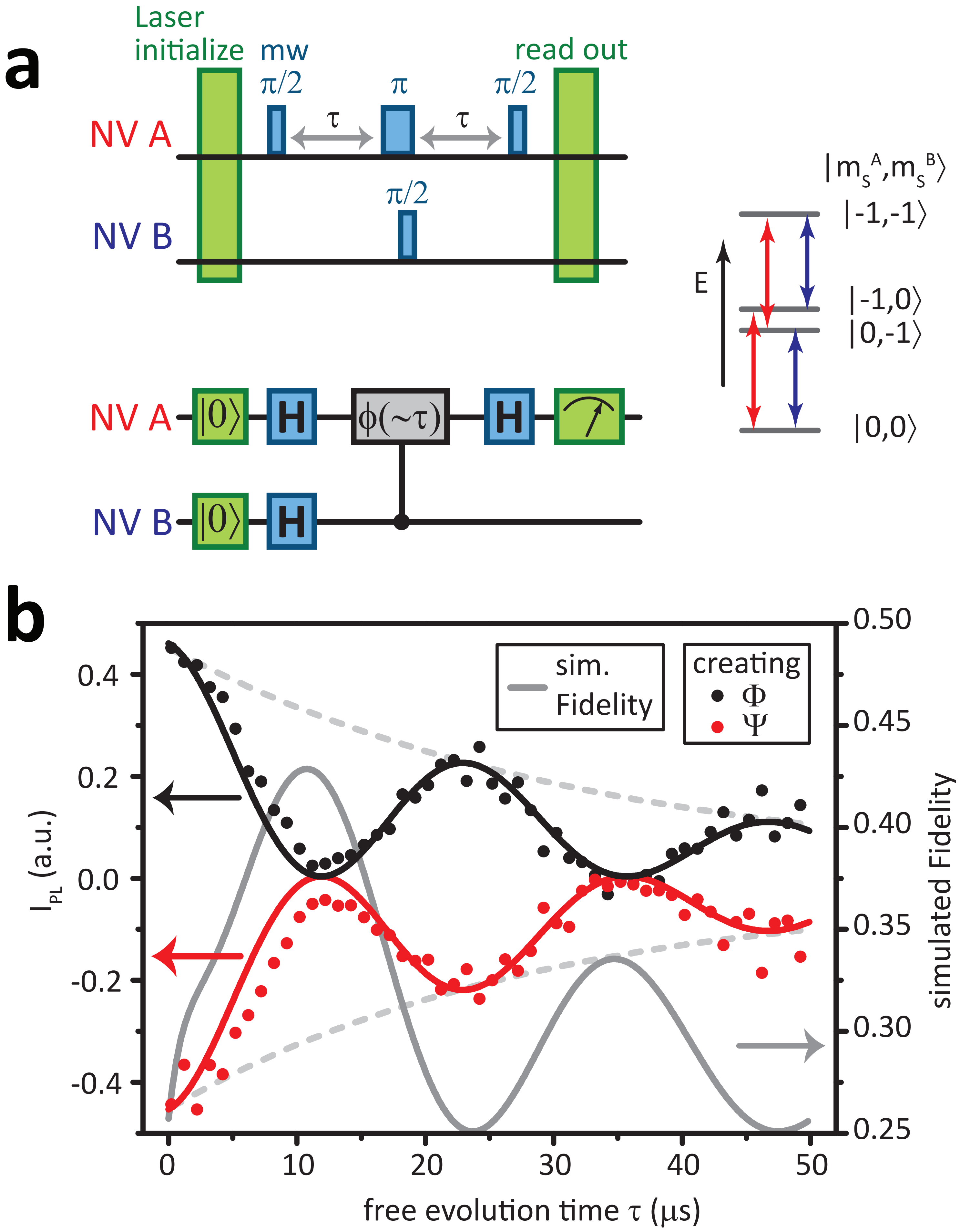

SUPPLEMENTARY METHODS AND DISCUSSION FOR

# Scalable quantum register based on coupled electron spins in diamond at room temperature

P. Neumann[1], R. Kolesov[1], B. Naydenov[1], J. Beck[1], F. Rempp[1], M. Steiner[1], V. Jacques[1], G. Balasubramanian[1], M.L. Markham[2], D.J. Twitchen[2], S. Pezzagna[3], J. Meijer[3], J. Twamley[4], F. Jelezko[1] and J. Wrachtrup[1]

[1] 3. Physikalisches Institut, Universität Stuttgart, Stuttgart, 70550, Germany

[2] Element Six Ltd., King's Ride Park, Ascot, Berkshire SL5 8BP, UK

[3] Central laboratory of ion beam and radionuclides, Ruhr Universität Bochum, 4480, Germany

[4] Center for Quantum Computer Technology, Macquarie University, Sydney 2109, Australia

* These authors contributed equally to this work

# ADDITIONAL METHODS

## Sample preparation:

The diamond sample is an ultrapure type IIa synthetic crystal produced by microwave plasma assisted chemical vapour deposition. The concentration of $^{12}C$ has been increased to 99.99% by using isotopically enriched methane in the growth process. NV centres have been created by implantation of $^{14}N^{3+}$ ions with energy of 13 MeV and subsequent annealing at 800°C for 2 hours. For such high implantation energy the nitrogen to NV conversion efficiency is better than previously reported owing to larger number of vacancies created (average NV creation efficiency 21%). In order to increase the probability to create pairs of NV centres at small separations, several arrays of implantation sites with varying numbers of ions per spot have been created.

## Fluorescence lifetime imaging (FLIM):

The usual confocal microscopy scan of an implantation area gives a first indication for the locations of NV-pair candidates. The fluorescence intensity exhibited from a small cluster of NVs in one confocal spot will be a multiple of the fluorescence intensity of a single centre from (double intensity in case of Figure S1a). From measurements of the second order correlation function this can then be verified (Fig. S1a). However, as the confocal contrast is too low to independently resolve the confocal spots (Ø ≈ 0.5 μm) of two NV centres separated by ≈10 nm, fluorescence lifetime imaging (FLIM) has been used to determine the exact relative position by optical means. This technique relies on the fact that the targeted emitters exhibit a different fluorescence lifetime. Instead of taking an image by recording the fluorescence intensity pixel by pixel FLIM also records fluorescence decay data per pixel. Short laser pulses are applied to bring the system to the excited state (Fig. S1b). The time delays of the subsequent photons are stored in a histogram. After the image is taken the histogram for every point can be fitted by e.g. a double exponential decay, with the two different decay rates for two different emitters. The amplitudes of these two exponentials are the pixel information for two additional images, one for each lifetime (Fig. S1b).

In case of the NV centres the fluorescence lifetime in bulk diamond is 11 ns for the $m_S^{A,B} = 0$ $\left(|0\rangle_{A,B}\right)$ spin sublevel and 7 ns for the $m_S^{A,B} = \pm 1$ $\left(|\pm 1\rangle_{A,B}\right)$ sublevels. Thus in order to have two emitters with two different fluorescence lifetimes we have to make sure that one centre is in $|0\rangle$ state while the other one is not. In the case of the two centres in Figure 1 their NV axes are different. Therefore a magnetic field of 700 Gauss applied parallel to the axis of one NV centre will initialize this one to $|0\rangle$ under optical illumination, while the other centre for which the magnetic field is misaligned, will be in a mixture of $|0\rangle, |\pm 1\rangle$. This ensures different fluorescence lifetimes for the two centres. A 2D correlation of these two images will result in a Gaussian like distribution with its centre being at the position of the displacement vector of the two centres. For the present case this shows a distance of 8±3 nm for the two centres in the plane of the diamond surface (Fig. 1d and Fig. S1c).

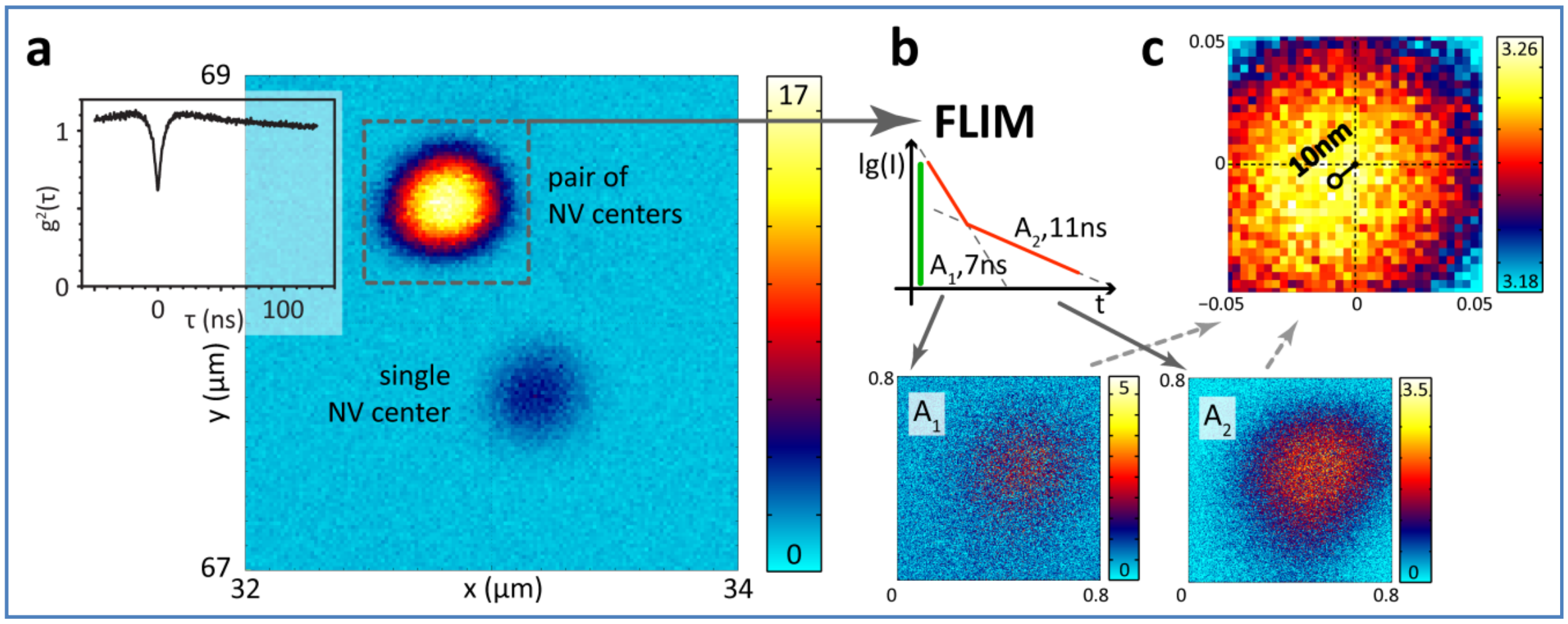

**Figure S 1 | Resolving a pair of NV centres**

**a,** Confocal microscopy image of a pair of NV centres and a single centre. Inset shows the fluorescence autocorrelation function $g^2(\tau)$. For $\tau$=0 it is ≈0.5 which proves the existence of two emitters. **b,** Fluorescence lifetime imaging (FLIM). The fluorescence decay curve (red line) after a picosecond excitation laser pulse (green line) for every pixel is fitted by a double exponential decay with amplitudes $A_1$ and $A_2$ and lifetimes of 7ns and 11ns. This results in two figures with the amplitudes $A_1$ and $A_2$ of the two exponentials. **c,** 2D correlation of the two figures $A_1$ and $A_2$ yields the separation of the two emitters.

## Ground state depletion microscopy of single NV centres:

Diffraction of light limits the resolving power of far-field imaging to about half of the wavelength of light. This limit first formulated by Abbe holds for only linear imaging and can be beaten in case when optical response of the system in nonlinear. Spectacular examples of such techniques such as STED, Palm and RESOLF were developed during last decade. In STED [S1] experiments the fluorescent object is illuminated by two optical beams. First the Gaussian beam excites fluorescence whereas the structured depletion beam having a minimum at the centre of excitation quenches fluorescence by inducing a stimulated transition between excited and ground state of the fluorescent marker. The key point of superresolution microscopy is a nonlinearity of the system response to the depletion beam intensity. When the depletion beam saturates the system, only the zero-intensity point remains unaffected. Hence the resolution limit is improved with a scaling factor proportional to the square root of the intensity of the excitation beam.

As described above the STED technique uses two laser beams which have to be overlapped with high precision. In principle the use of a single beam having a zero-intensity point also can provide a superresolution image. In such simplified scheme the structured illumination beam saturates the absorption of the marker [S2,3]. Therefore the zero intensity point provides a subdiffraction negative spot in the fluorescence image. Examples of ground state depletion microscopy of single colour centres are given in Figure S2. The half width of the image for such nonlinear technique is given by $R = r_0\sqrt{\Gamma/P_0}$, where $r_0$ is the half width of diffraction limited illumination spot, $\Gamma$ is radiative decay rate and $P_0$ is optical pumping rate. Owing to excitation power limitations we were able to reach 20 nm spot size in our set-up. Such resolution allows identifying the centres that might be suitable for further investigation using spin resonance.

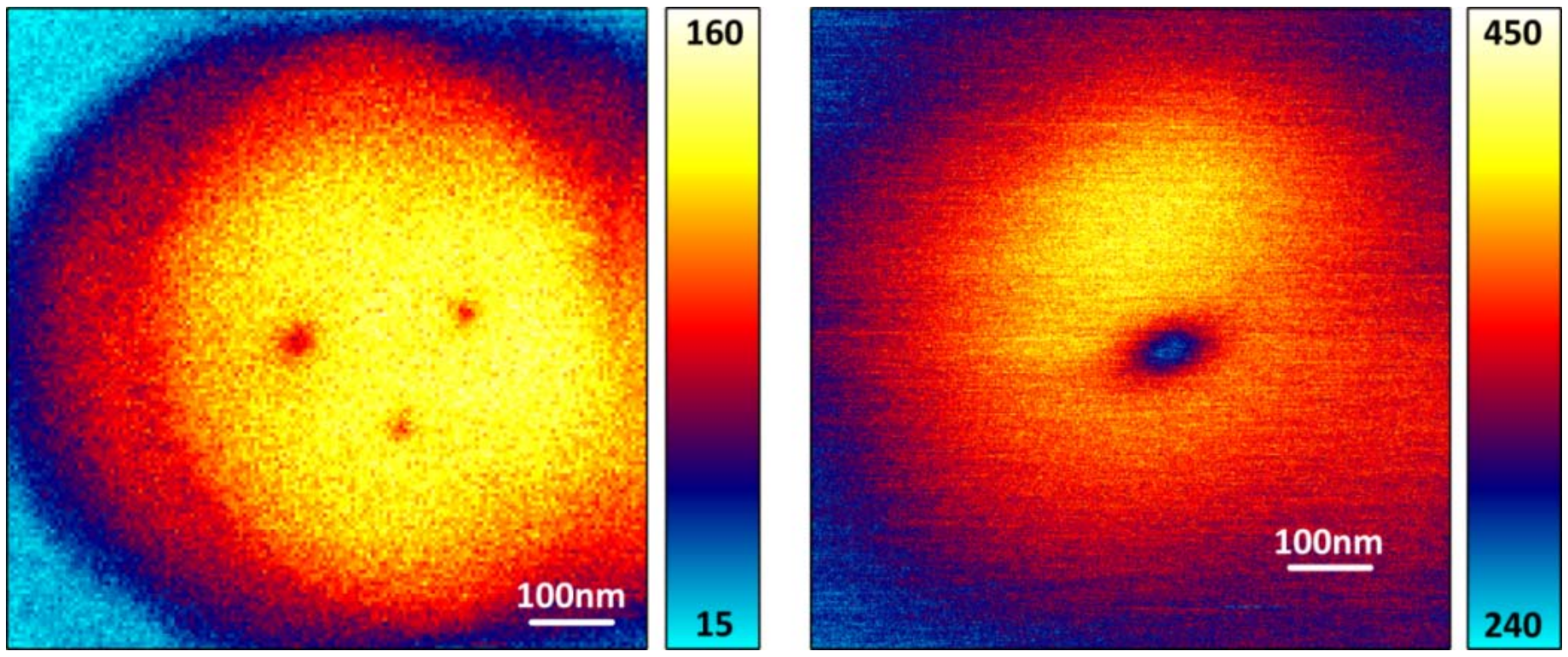

**Figure S 2 | Ground state depletion microscopy.**

Ground state depletion microscopy images of 3 colour centres (left) in close vicinity. Right graph represents the image of the two defects studied in this paper. Note that owing to small separation their individual positions were not resolvable by this technique.

## Hamiltonian, Energy levels and eigenstates:

The Hamiltonian of two magnetically coupled spins is:

$$H = \frac{\mu_0 g^2 \mu_B^2}{4\pi r^3}\left[\hat{S}_{NV\,A}\cdot\hat{S}_{NV\,B} - 3\left(\hat{S}_{NV\,A}\cdot\vec{r}\right)\left(\hat{S}_{NV\,B}\cdot\vec{r}\right)\right] + \sum_{i=NV\,A}^{NV\,B}\left(\hat{S}_i\underline{\underline{D}}\hat{S}_i + \mu_B g_e \underline{B}\hat{S}_i + \hat{S}_i\underline{\underline{A}}\hat{I}_i\right) \tag{S1}$$

This first part is the magnetic dipole-dipole coupling of the two centre's spins and the second one is the sum of the single centre spin Hamiltonians.

Firstly, we want to point to some specialties of the eigenstates and energy levels that result from the fact that the spin has S=1 with a zero field splitting of $H_{zfs}$=2.87 GHz. The magnetic field strengths used here yield Zeeman energies ($H_{Zeeman}$), much smaller than $H_{zfs}$. In addition there is always a component of the field, $B_{\|}$, parallel to the NV axes. So even in the case of an applied perpendicular component $B_{\perp}$, the $|\pm1\rangle$ levels almost keep their magnetic moment to first order. However now the $|0\rangle$ level acquires a magnetic moment perpendicular to the quantization axis which grows linearly with $B_{\perp}$ (Fig. S3a).

This has an influence on the magnetic dipole-dipole interaction energies for the different eigenstates. While the energies of states without $|0\rangle$ remain almost unchanged, those states with $|0\rangle$ get shifted when taking into account this perturbation. The strength of this shift depends on the direction and strength of $B_{\perp}$. Larger values of $B_{\perp}$ increases the induced magnetic dipole of the $|0\rangle$ level, while the direction of this magnetic field component changes the direction of the dipole. Finally, the direction and strength of this induced dipole determines the interaction energy with the other spin (see respective term in Equation (S1)).

This shift is observed in the Ramsey fringes (Fig. 2b, S3b) as well as in the echo based experiments (Fig. S5a) and it is the basis for determining the relative position of the two centres using electron spin resonance (Fig. S5c).

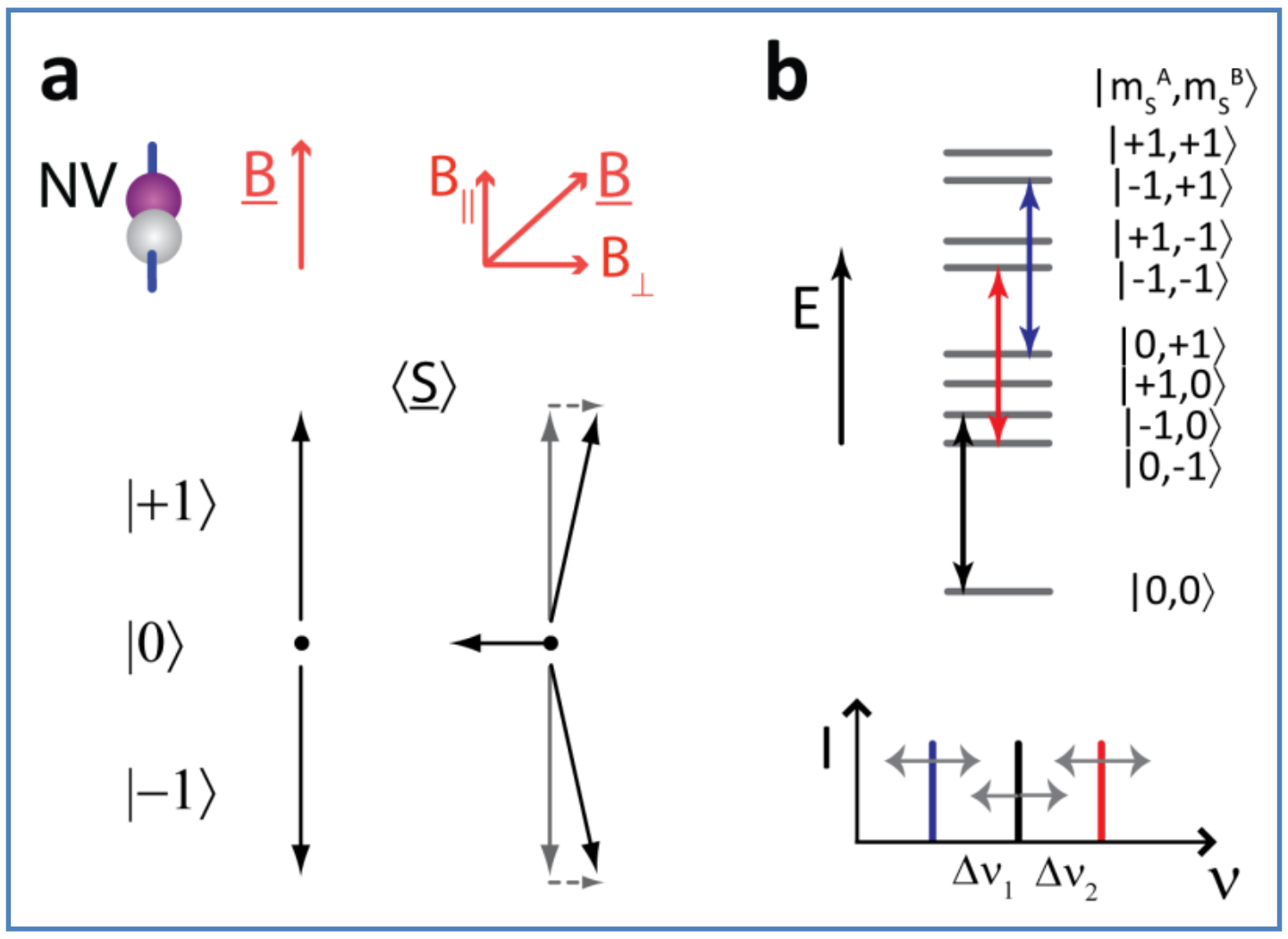

**Figure S 3 | Spin states depending on magnetic field**

**a,** Expectation value of the spin vector for the three eigenstates. For a magnetic field along the NV axis nothing changes. For a tilted field every eigenstate acquires a small ⊥-component of spin orientation. **b,** Energy levels of the coupled NV –NV system. The three shown transition frequencies differ by the coupling frequency. For tilted magnetic fields $\Delta\nu_1$ and $\Delta\nu_2$ change independently.

## Spin control using pulse techniques

## Free induction decay (Ramsey fringes):

To acquire a usual optically detected magnetic resonance (ODMR) spectrum a NV centre is continuously illuminated with green laser light and the frequency of the microwaves guided to the NV is swept while the fluorescence intensity $I_{PL}$ is recorded (Fig. S4a). The obtained spectrum is broadened by laser and microwave power. Using the obtained resonance frequencies coherently driven Rabi oscillations can be performed (Fig. 3b, c). These yield the length of transition selective π/2- and π-pulses. In order to obtain a detailed ESR spectrum of the two-NV-system free from power broadening, the free induction decay can be recorded. Both NV centres are initialized to $|0\rangle$ by a laser pulse. NV $A$ is now brought into a coherent superposition between two of its eigenstates $\psi_A(0) = (|0\rangle + |-1\rangle)/\sqrt{2}$ by a π/2-pulse which is slightly detuned from the observed ESR-line. The phase of this superposition now evolves with the detuning frequency ν during a free evolution time interval τ $\psi_A(\tau) = (|0\rangle + e^{i2\pi\nu\tau}|-1\rangle)/\sqrt{2}$. A second π/2-pulse converts phase into a spin population difference which is read out by another laser pulse. This fluorescence signal $I_{PL}$ as a function of τ shows an oscillation with frequency ν ($I_{PL} \sim \langle 0|\psi|0\rangle = cos^2(\pi\nu\tau)$, called Ramsey fringes) which decays after a time $T_2^*$. The Fourier transformation of the signal yields the exact relative spectral position of the observed line with a width proportional to $1/T_2^*$ (Fig. 2b).

Because the two NV centres are coupled by a magnetic dipole-dipole interaction, the energy and thus the line position of NV $A$ depends on the spin state of NV $B$. In the picture of the evolving spin superposition state of NV $A$, its oscillation frequency ν is increased or reduced by $\Delta\nu_{1,2}$ due to the additional magnetic field that NV $B$ exerts on it (Fig. S3b, 2b). Through resonant π-pulses, NV $B$ can be prepared in one of the spin states $|0\rangle_B, |\pm1\rangle_B$ shifting the spectral line position of NV *A* for each state (Fig. 2b). This reveals a coupling frequency of about Δν≈40 kHz. Note that environmental changes might also lead to changes in line positions. To exclude this all three measurements for the different spin states of NV *B* were recorded simultaneously. The splitting of the three lines doesn't have to be equal due to the reasons mentioned above.

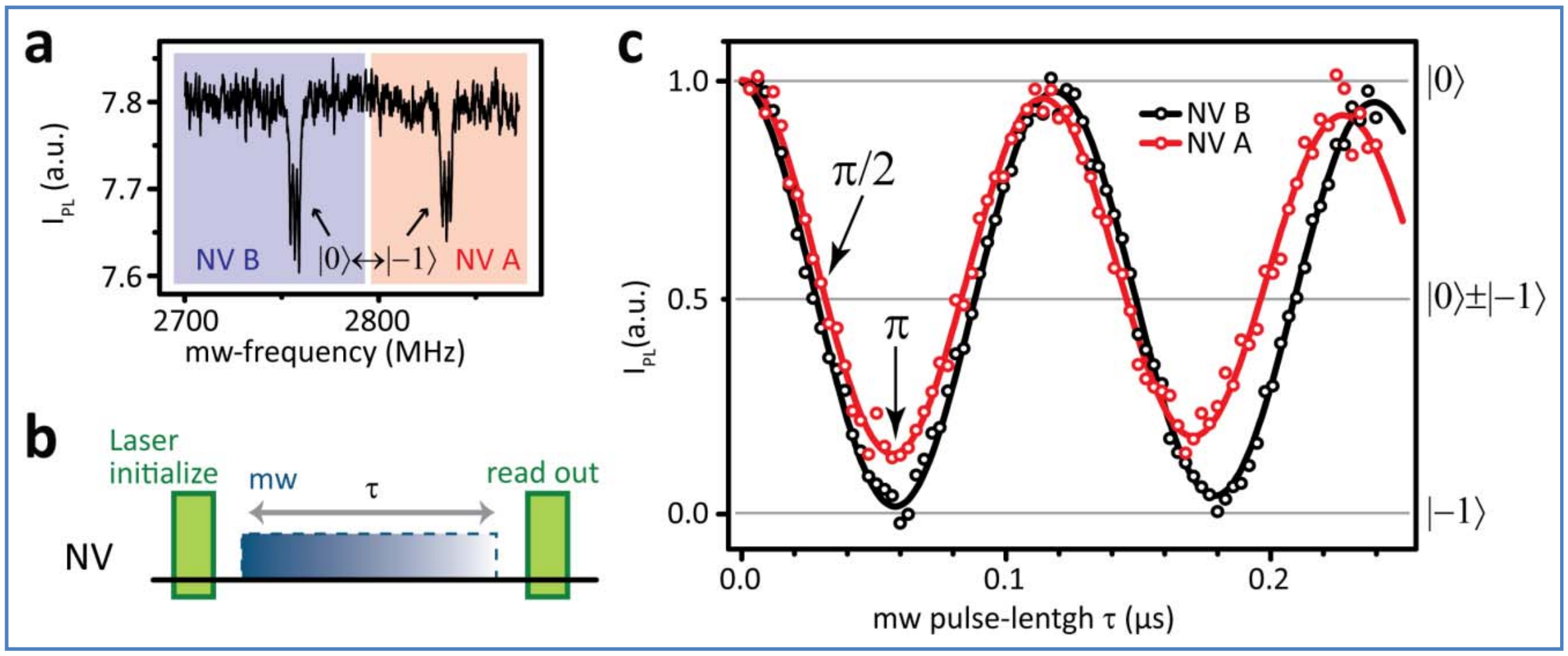

**Figure S 4 | Single electron spin Rabi oscillations**

**a,** ODMR spectrum of a NV centre *A* and *B* showing their $|\mathbf{0}\rangle \leftrightarrow |-\mathbf{1}\rangle$ spin transitions. **b,** Sequence to perform a Rabi nutation. The laser initializes the NV centre into the $|\mathbf{0}\rangle$ state and the resonant microwave pulse is steadily increased. The final laser pulse reads out what is left in the $|\mathbf{0}\rangle$ state for every microwave pulse length τ. **c,** Oscillating fluorescence intensity reveals the coherent Rabi nutation of a single electron spin (NV *A* - red line, NV *B* - black line). The corresponding quantum states are marked on the right side. Pulse lengths for π/2- and π-pulses can be obtained.

## Echo-based detection of spin-spin coupling

In contrast to the Ramsey fringes, echo based techniques can compensate any slow magnetic field fluctuations created by the nuclear spin bath for instance. These echo protocols effectively increase the coherence time from $T_2^*$ to $T_2$ (i.e. from 10 µs to 110 µs in case of spin *A*).

The electron-electron double resonance (DEER) pulse sequence is a Hahn echo on spin *A*, combined with a flip of spin *B* at varying time instants (Fig. 3a, S5a). In a Hahn echo sequence, a laser pulse initializes the NV centre to $|0\rangle$, then a $\pi/2$-pulse creates a coherent superposition of the centre's spin $\psi_A(0) = (|0\rangle + |-1\rangle)/\sqrt{2}$. The phase then evolves during a time interval τ, $\psi_A(\tau) = (|0\rangle + e^{i2\pi\nu\tau}|-1\rangle)/\sqrt{2}$. A π-pulse then inverts the phase, $\psi_A(\tau) = (|0\rangle + e^{-i2\pi\nu\tau}|-1\rangle)/\sqrt{2}$, which begins to re-evolve towards the situation after the first $\pi/2$-pulse. After another time interval τ, a second π/2-pulse pulse converts the phase back to spin population difference which is read out by a laser pulse. This sequence can refocus phase shifts that occur on a timescale slower than 2τ. The echo signal as a function of 2τ decays after a phase memory time $T_2$. However, if the phase is shifted e.g. during the re-evolution period, it will not be refocused. To measure the coupling between the two NV centres, a Hahn echo sequence is applied to NV *A* with a fixed time τ while NV *B* is flipped by a π-pulse. This spin flip of NV *B* changes the magnetic field felt by NV *A* by $\delta B$ and therefore the accumulated phase Δφ at the end of the echo sequence. Varying the time instant T of the spin flip of NV *B* will alter this phase $\Delta\varphi = \gamma \cdot \delta B \cdot T = \Delta\nu_{1,2} \cdot T$ which is accumulated during the sequence. Therefore, the echo signal as a function of the temporal position of the π-pulse on NV *B* shows a modulation which oscillates with exactly the coupling frequency of the NV centre pair (Fig. 3, S5a). To ensure that the observed signal is not an artefact, this sequence has also been carried out with the π-pulse on NV *B* being off resonant. In this case no oscillation was visible and NV *B* was not flipped (Fig. S5a). Higher oscillation frequencies which are important for faster quantum gates can be achieved by exploiting the triplet character of the NV centre's spin. When spin *B* is flipped from $|-1\rangle_B$ to $|+1\rangle_B$ the magnetic field change that spin *A* feels nearly doubles (Fig. S3b) and so does the oscillation frequency ($\Delta\nu_1$+$\Delta\nu_2$, Fig. S5b, red line). In addition the sensitivity of spin A can be enhanced by generation the superposition of $|-1\rangle_A$ and $|+1\rangle_A$, $\psi_A(0) = (|-1\rangle + |+1\rangle)/\sqrt{2}$, which again doubles the oscillation frequency (Fig. S5b, grey line).

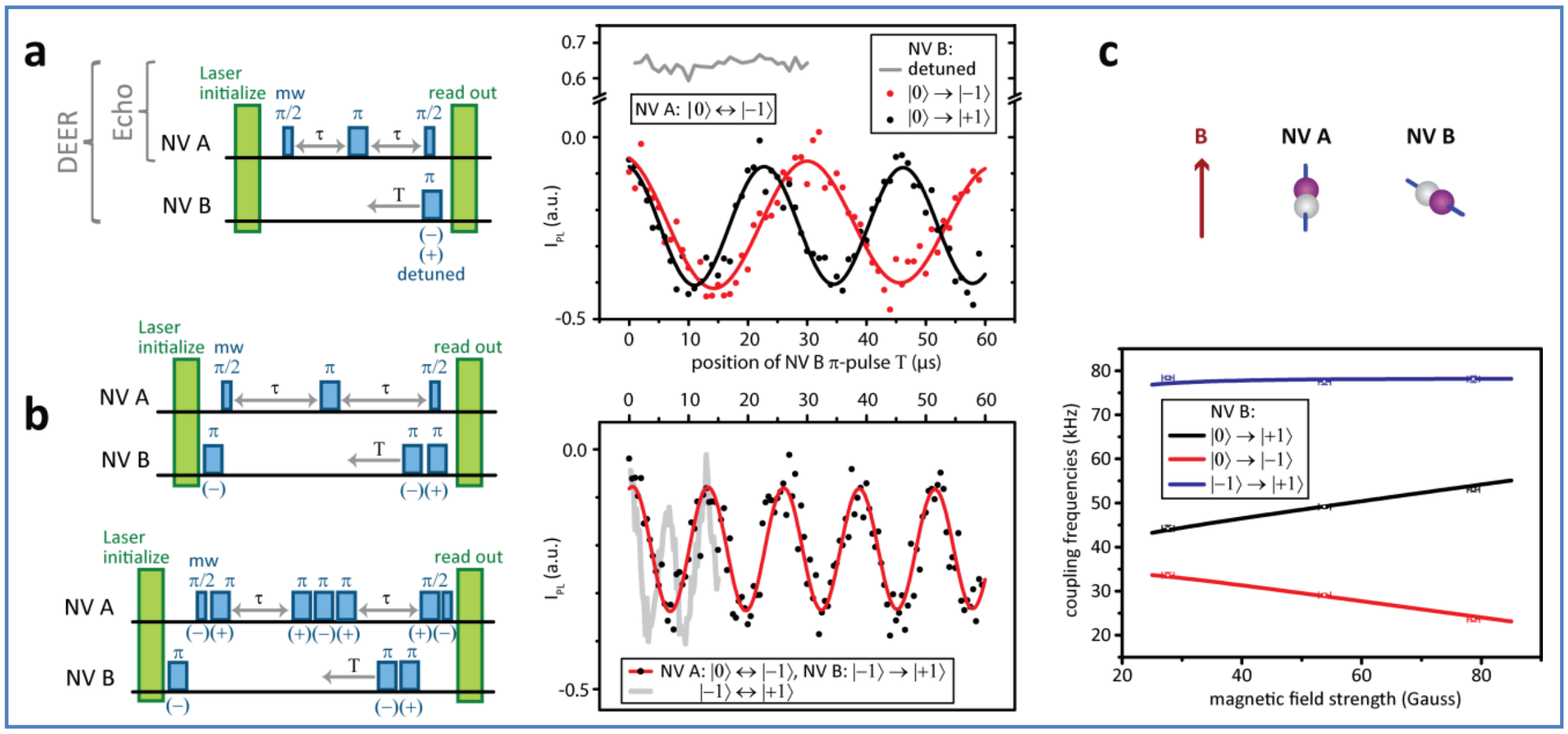

**Figure S 5 | Double electron-electron resonance (DEER).**

**a,** Pulse sequence to detect coupling strength of the two centres by DEER. An echo is performed on a transition of NV *A* with fixed τ. During the second interval τ a π pulse NV *B* transition $|\mathbf{0}\rangle_{\boldsymbol{B}} \leftrightarrow |-\mathbf{1}\rangle_{\boldsymbol{B}}$ (-) or $|\mathbf{0}\rangle_{\boldsymbol{B}} \leftrightarrow |+\mathbf{1}\rangle_{\boldsymbol{B}}$ (+) is swept (T). The resulting oscillations with frequencies $\Delta\nu_1$ and $\Delta\nu_2$ are shown. If the □ pulse on NV *B* is detuned its spin is not flipped (→higher fluorescence level) and no oscillation is visible (grey line). **b,** same experiments now performed using also transition $|-\mathbf{1}\rangle_{\boldsymbol{B}} \leftrightarrow |+\mathbf{1}\rangle_{\boldsymbol{B}}$ for higher frequency and thus quantum gate speed. If NV *B* is flipped from $|-\mathbf{1}\rangle_{\boldsymbol{B}}$ to $|+\mathbf{1}\rangle_{\boldsymbol{B}}$ (upper sequence) the frequency is almost doubled ($\Delta\nu_1+\Delta\nu_2$ red line). If, in addition, the coherence during the echo on NV *A* is stored on transition $|-\mathbf{1}\rangle_{\boldsymbol{A}} \leftrightarrow |+\mathbf{1}\rangle_{\boldsymbol{A}}$ (lower sequence) the frequency is again doubled (grey line). **c,** Coupling frequencies of DEER measurements depend on magnetic field. For the shown arrangement of magnetic field parallel to NV *A* increasing strength leads to decrease of $\Delta\nu_2$ (red data) and an increase of $\Delta\nu_1$ (black data). The sum remains almost unchanged (blue data). The data is fitted by adjusting the relative position of the two centres and diagonalising the Hamiltonian. Actually, taking several series of DEER frequencies like this one for different fields enables finding the relative position of the two centres.

## Structure of the spin pair

The exact relative position and orientation of the two NV centres can be inferred from the results of different series of DEER measurements. As explained above a magnetic field component perpendicular to the NV axis induces a magnetic dipole in the $|0\rangle$ state (Fig. S3a). Thus, by changing the strength and direction of the magnetic field the energy splitting induced by the spin-spin interaction is changed (Fig. S3b). This energy splitting can be determined by measuring various DEER frequencies. That is: performing a Hahn echo on NV $A$ for transitions $|0\rangle_A \leftrightarrow |-1\rangle_A$, and $|0\rangle_A \leftrightarrow |+1\rangle_A$ and for each echo perform the transition on NV $B$ from $|0\rangle_B \rightarrow |-1\rangle_B$ and $|0\rangle_B \rightarrow |+1\rangle_B$ at time T. ($|-1\rangle_B \rightarrow |+1\rangle_B$ would be for redundancy).

One DEER series has been measured for the magnetic field aligned along NV $A$ and for 3 different field values (Fig. S5c). A second series is taken for a field direction almost along a third possible NV axis, which means the field is strongly misaligned for both defect centres. The measured values are now compared to the corresponding values of the diagonalised Hamiltonian. Distance and orientation of the two centres are altered to obtain the best fit. The uncertainty region of the relative position has a volume which is slightly smaller than a unit cell (Fig. 3c). And depending on the two possible lattice configurations there are 6 or 5 lattice positions for the other electron spin. Note that the Hamiltonian assumes a point dipole. Actually, the electron spin density is spread mainly over the 3 dangling bonds of the nearest neighbour carbon atoms around the vacancy. That is quite localized for a distance of ≈10 nm but the localization accuracy comes close to this length scale. If one wants to localize the two NV centres even more precisely it might be necessary to take the spread of the spin wave-function into account.

## Entanglement generation between spins

We developed a pulse sequence to act as an entangling gate which creates Bell states between the two electron spins (description in text). Note that for the present NV pair the entanglement generation does not show high fidelity due to the short coherence lifetime $T_2$ of NV *B*. We simulated the effect of the pulse sequence on the system by using the corresponding unitary transforms for the pulses and the free precession of the system. Additionally, we applied exponential dephasing to the state during the free evolution of the system. Results are shown in Figure 4b.

We used the same technique to estimate the possible fidelities for longer $T_2$ of NV $B$. The fidelity $F = \langle \Theta | \Psi | \Theta \rangle$ is used to determine how close a given state Θ is to a target state $|\Psi\rangle$. In our case $|\Psi\rangle$ is a Bell state. If we assume a $T_2$ value of 200 μs for both spins the fidelity of the created Bell states should reach more than 90%. And if we even assume those $T_2$ values measured previously in ultrapure isotopically enriched diamond samples, namely more than 1 ms, the fidelity becomes close to 1 (>99%) (Fig. S6).

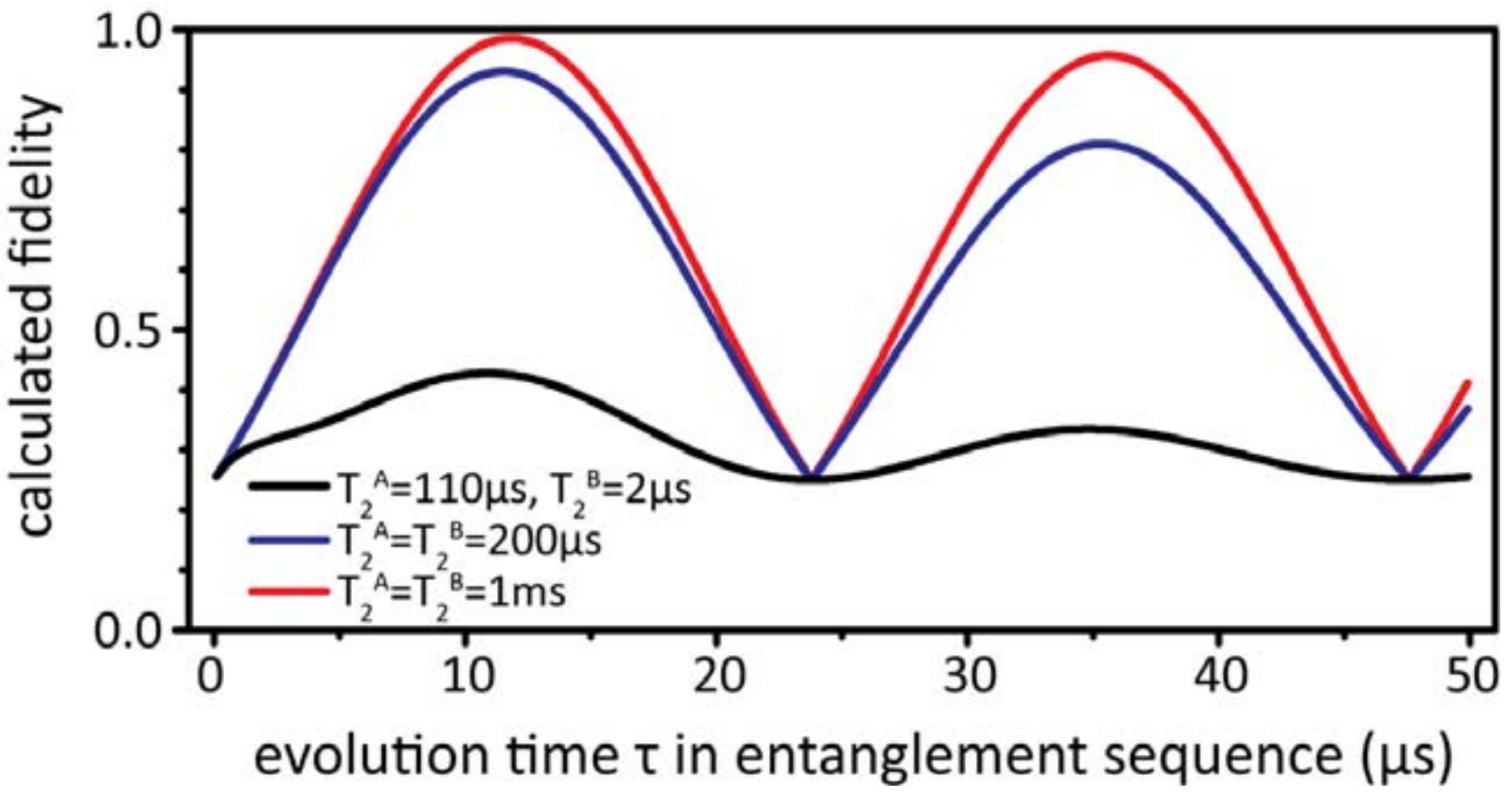

**Figure S 6 | Fidelity of entangled states**

Fidelity of the created Bell state for the shown entanglement sequence (Fig. 4, in the paper) calculated for different phase memory times $T_2$ of both centres. For times of 1ms which is reasonable in this sample the fidelity reaches 1.

# DISCUSSION

## Coherence time of implanted spin pair.

The origin of the drastic difference in the coherence properties of the two closely spaced NV defects can be attributed to the fabrication process. We have found that the zero field splitting parameters (D and E) for two implanted centres forming a spin dimer differ significantly from isolated spins. Since nitrogen–vacancy colour centres relevant to these experiments are negatively charged, we expect that Coulomb interaction will affect the ESR spectra of spin dimers (corresponding field strength 2.7 MV/m). However it is known that electric fields perpendicular to the nitrogen-vacancy axis leads to an ESR line splitting which scales as 0.17 MHz/(MV/m) [S4]. Therefore we expect that the electrostatic interactions shift the energy levels of the two colour centres by only $E_A$=0.21 MHz and $E_B$=0.42 MHz (calculated from relative position and orientation). Surprisingly our measured data show much higher values for such shifts: $E_A$=2.3 MHz and $E_B$=5 MHz. We thus conclude that these larger shifts are most likely due to localized strain caused by damage during the implantation process. These strained and damaged regions can have a dynamic character and will lead to localized sources of decoherence. Further experiments employing annealing of crystals at high temperature with the full reconstruction of the diamond lattice will be carried out in order to improve the coherence times of implanted defect pairs in future.